# Gating of single molecule junction conductance by charge transfer complex formation.

Andrea Vezzoli[a], Iain Grace[b], Carly Brooke[a], Kun Wang[c], Colin J. Lambert[b] Bingqian Xu[c], Richard J. Nichols[a] and Simon J. Higgins[a*]

[a.] Department of Chemistry, Donnan and Robert Robinson Laboratories, University of Liverpool, Liverpool L69 7ZD, United Kingdom.
[b.] Department of Physics, Lancaster University, Lancaster LA1 4YB, United Kingdom.
[c.] College of Engineering & NanoSEC, University of Georgia, 220 Riverbend Road, Athens, Georgia GA 30602, U.S.A.

The solid-state structures of organic charge transfer (CT) salts are critical in determining their mode of charge transport, and hence their unusual electrical properties, which range from semiconducting through metallic to superconducting. In contrast, using both theory and experiment, we show here that the conductance of metal | *single molecule* | metal junctions involving aromatic donor moieties (dialkylterthiophene, dialkylbenzene) increase by over an order of magnitude upon formation of charge transfer (CT) complexes with tetracyanoethylene (TCNE). This enhancement occurs because CT complex formation creates a new resonance in the transmission function, close to the metal contact Fermi energy, that is a signal of room-temperature quantum interference.

## Introduction

The discovery that treatment of the aromatic hydrocarbon perylene with bromine resulted in the formation of a black, crystalline and surprisingly conductive ($10^{-2}$ S cm$^{-1}$) organic material[1] initiated the systematic study of charge transfer complexes. With the development of strong molecular acceptors such as tetracyanoquinodimethane (TCNQ)[2] and tetracyanoethylene (TCNE),[3] and of donors such as tetrathiafulvalene (TTF),[4] this gave rise to ground-breaking discoveries, such as the first molecular metal (the charge transfer salt TTF.TCNQ[5]) and the first molecular superconductor ([TMTSF]$_2$PF$_6$; TMTSF = tetramethyl-tetraselenofulvalene).[6,7] Following the syntheses of regioregular poly-3-alkylthiophenes[8,9] and the discovery that thin films of these materials can have good semiconductor properties,[10] oligo- and polythiophene derivatives have become widely-studied materials in organic electronics; their structures have been tuned to enable them to exhibit very high field effect mobilities in organic transistors, and they are also good light absorbers and hole-transport materials in heterojunction organic photovoltaic devices.[11,12] The solid-state structures of charge transfer salts are crucial in determining charge transport through the structures, and hence their unusual electrical properties. In contrast, we show here that the conductance of metal | *single molecule* | metal junctions involving either a terthiophene or a simple 1,4- phenylene moiety increase by over an order of magnitude upon formation of their charge transfer complexes with TCNE, and that this large conductance enhancement is due to the creation of a new quantum resonance in the transmission function of the junction upon complex formation that is close to the contact Fermi energy.

The development of reliable techniques for fabricating and electrically characterizing metal | molecule | metal junctions has stimulated a revival in interest in molecular electronics, partly driven by the realization that the continued downsizing of individual components in silicon chips could mean that individual molecules might eventually play some role in electronic devices. Early studies of this type often addressed the question of the most efficient electrical transmission over long distances, by optimising such factors as the nature of the metal-molecule contacts and the degree of conjugation of the backbone.[13-15] The possibilities offered by external control over molecular conductance, for instance using a third (gate) electrode,[16] or electrochemical potential,[17-20] or by irradiation with light[21-23] have clear implications for future device applications. However, rather than seeking molecular analogues of classical inorganic devices, the pursuit of phenomena unique to nanoscale molecular junctions may be of more interest. For example, cross-conjugated molecules offer the possibility for large thermoelectric effects based upon destructive interference,[24,25] and by using specific molecule…molecule interactions (host-guest interactions[26]) it may be possible to use molecular junctions as uniquely sensitive sensor devices.

We previously showed that the behaviour of single–molecule junction conductance as a function of molecular length for the oligothiophenes **1a-d** (Figure 1) depended upon the presence or absence of water.[27] In the presence of water, junctions involving molecules **1a-d** all had similar conductances (*ca.* $10^{-5} G_0$ where $G_0$ is the quantum unit of conductance, $2e^2/h$, or $7.748 \times 10^{-5}$ S). In the absence of water, the experimental conductances fell exponentially as a function of length, as had been predicted by non-equilibrium Green's function-based (NEGF) transport calculations.[27] Water molecules have a weak dipolar interaction with the thiophene π-bonding orbitals; for the longer, more conjugated examples with transport resonances closer to the Fermi energy ($E_F$) of the contacts, this results in a small shift in energy of these resonances towards $E_F$, sufficient to offset the decay of conductance with length.

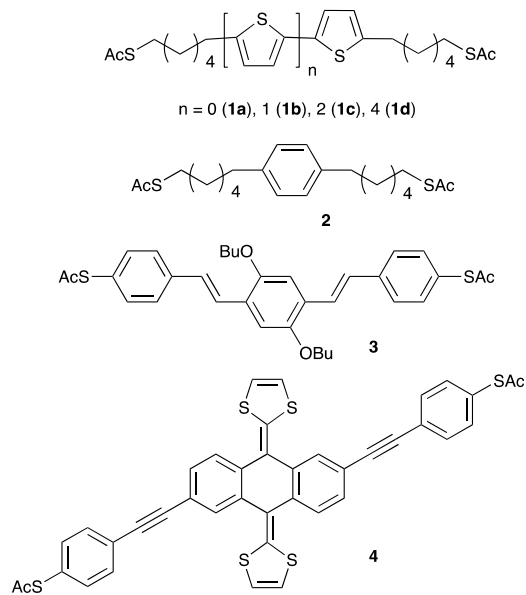

Figure 1 Structures of molecules discussed in the text.

Since simple oligothiophenes (2,2′-bithiophene, 2,2′:5′,2′′-terthiophene) form charge transfer complexes with acceptors such as TCNE, TCNQ and its derivatives, and $I_2$,[28-30] we reasoned that this could have an even stronger effect upon junction conductance. Accordingly, we have now explored the effect of charge transfer complexation using TCNE upon the junction conductance of representative molecules, **1c** and **2**, using both experimental and theoretical techniques. We find a substantial increase of conductance on charge transfer complex formation, and using DFT and NEGF transport calculations, we are able to attribute this to the appearance of a new resonance (a Fano resonance) near to $E_F$ in the transmission function as a result of charge transfer complex formation, and we report these results here.

While this work was in progress, two other reports of the effect of charge transfer complexation upon molecular junction conductance appeared. Using a version of the 'matrix isolation' method described earlier by Cui et al.,[31] Del Re et al. found a significant decrease (ca. 5-fold or 20-fold, depending upon how the data are interpreted) in the conductance of Au-molecule-Au junctions with oligophenylenevinylene **3** (Figure 1) upon prior exposure to 1,4–dinitrobenzene, attributing this to changes in the HOMO energy as a result of a charge transfer interaction.[32] However, in this case a wide spread of conductance values is evident. In contrast, the cruciform extended tetrathiafulvalene derivative **4** showed a conductance increase upon exposure to the highly electron-deficient tetrafluorotetracyanoquino-dimethane ($F_4$TCNQ), although in this case, the conductance of the uncomplexed **4** was too small to measure (i.e. <$10^{-7}$ $G_0$), so that the factor by which the conductance increased could not be determined. In addition there were complications caused apparently by different binding modes of the molecules in the junctions.[33] Moreover, spectroscopic data indicated that the powerful oxidant $F_4$TCNQ predominantly fully oxidises **4**, hence switching it from its cross-conjugated, neutral form to a fully-conjugated, oxidised form. This would certainly be expected to increase the conductance.[34] We chose the weaker acceptor TCNE because (i) redox chemistry with the donor molecules **1c**, and particularly **2**, is less likely, and (ii) TCNE is small enough to form a charge transfer complex with the single arene ring of **2**.

## Results and discussion

### Experimental results

Terthiophene **1c** combines ease of synthesis (see ESI†) with sufficiently extended conjugation to facilitate charge transfer complex formation. We therefore first investigated the interaction of **1c** with TCNE in $CH_2Cl_2$ solution.

Treatment of 2 mM **1c** with 2 eq. of TCNE in $CH_2Cl_2$ resulted in the formation of a single new band in the visible spectrum at 11,800 $cm^{-1}$ ($\varepsilon$ 40 $dm^3mol^{-1}cm^{-1}$), due to intermolecular donor-acceptor charge transfer (CT) complex formation (ESI, Figure S11). This is at lower energy than for the TCNE complex of 5,5'-dimethyl-2,2'-bithiophene (15,400 $cm^{-1}$),[30] or of terthiophene itself (13,070 $cm^{-1}$),[35] as expected since **1c** is more conjugated than 5,5'-dimethyl-2,2'-bithiophene, and is more electron-rich than unfunctionalised terthiophene. We were unable to measure the equilibrium constant for CT complex formation $K_{DA}$ using the usual literature procedure[36] because of the overlap of the CT band with the terthiophene $\pi-\pi^*$ absorption band, but $K_{DA}$ for 5,5'-dimethyl-2,2'-bithiophene is 4.4 (in $CHCl_3$)[30] and $K_{DA}$ for **1c** should be larger than this.

We then employed the STM break junction technique[37] to measure the electrical properties of metal | molecule | metal junctions with **1c** in the presence and absence of TCNE. Briefly, a Au STM tip was brought close to the surface of a Au(111) substrate, in the presence of either **1c** or of the **1c**:TCNE complex. The feedback loop was disengaged and the tip was pushed into contact with the substrate. It was then retracted while the tunnelling current was monitored. A fresh Au-Au junction is formed, and on retraction this thins down to a single atom (point contact), which is finally broken upon further withdrawal. This process results in peaks in a conductance histogram at multiples of the quantum unit of conductance $G_0$ ($2e^2/h$; 7.75 × $10^{-5}$ S) owing to atomic rearrangements in the junction. When the point contact breaks, if no molecule is trapped in the resulting break junction then the final step down of $G_0$ is followed by a sharp exponential decay as the tip continues to retract, but if a molecule (or molecules) binds to both gold contacts, then subsequent additional plateau(s) are seen in the current-distance plot at conductances << $G_0$, corresponding to tunnelling conductance through the molecule. Eventually, as retraction continues, the metal | molecule | metal junction breaks down whereupon the current rapidly falls to a very low value consistent with tunnelling through space once more.

The highly variable nature of possible gold-thiol interactions, together with the conformational flexibility of the molecules and the correspondingly different conductance values, makes a statistical analysis of many such experiments essential. Accordingly, the procedure was repeated many times to obtain a statistical picture of junction formation and electrical properties. Typically, 5000 such traces were employed for each set of experimental conditions, and the traces were binned to produce one-dimensional histogram plots of log(conductance) *vs.* frequency of occurrence (see Experimental, and ESI Figures S5 and S6 for example traces). Importantly, no trace selection was employed in the construction of the conductance histograms. Figure 2 shows the results of such an experiment for molecule **1c** alone, and for the **1c**:TCNE complex, using a solution-based technique; a similar technique using a pre-formed self-assembled monolayer (*q.v.*) was also employed for comparison and the data for **1c** and the **1c**:TCNE complex are shown in Figures S7 and S8.

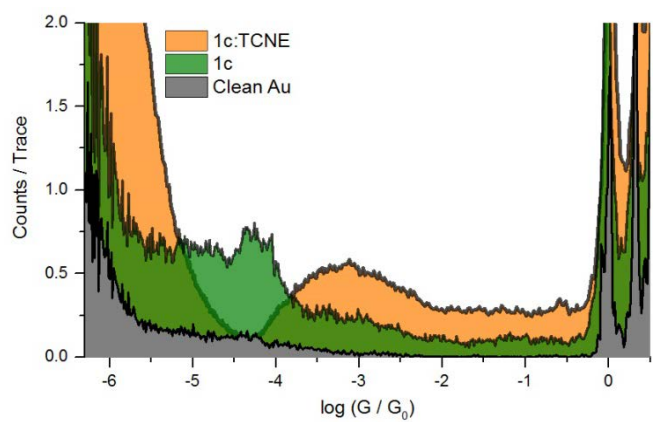

Figure 2 Logarithmic conductance histogram plots for molecule 1c (1 mM solution in TCB; green), and for 1c recorded in the presence of 10 mM TCNE in TCB (orange). Bias voltage 0.1 V in both cases. The results for a blank experiment with no molecules in solution are also shown (grey).

It can be seen that, in addition to the sharp peaks at multiples of $G_0$ owing to atomic rearrangements of the gold junction, there is a single very broad peak for the **1c**:TCNE complex, centred at a conductance of *ca.* $10^{-3}G_0$, or 77 nS, and that **1c** alone under these conditions shows a conductance peak centred at $10^{-4.3}G_0$, or 3.9 nS, a factor of 20 smaller. The large number of counts/trace at low conductance values for the complex may be a result of a Faradaic current due to the relatively large concentration of electroactive TCNE in this solution-based technique.

In control experiments, we first collected STM-BJ data on a bare gold slide in the presence of 1 mM TCNE in TCB to see whether any junction formation with TCNE itself occurred; no traces from $10^{-5.5}$ - $G_0$ (the limits of the solution method) consistent with junction formation were seen under any conditions tried. Additionally, we used a related non-contact STM method (the *I(s)* technique; see ESI) that has a lower minimum current limit, to examine down to $10^{-7}$ $G_0$; again, no current plateaus consistent with junction formation were observed. Next, we determined the conductance of junctions with 1,10-decanedithiol in TCB, in the presence and absence of 10 mM TCNE. The conductance peaks observed agreed with literature values and did not change significantly in the presence of TCNE (Figure S10; Table S1). This confirms that it is the formation of the charge transfer complex with the terthiophene unit of **1c** that is responsible for the conductance increase, and not either an interaction of the TCNE with the Au surface, and/or the Au-S contacts, as such effects would be expected likewise to affect the conductance of 1,10-decanedithiol junctions.

We have already shown that, in contrast to **1c**, the conductances of junctions involving molecule **2** are not affected by the presence or absence of water[38, 39], but it is known that alkylbenzenes do form CT complexes with acceptors like TCNE (indeed, they were among the first TCNE CT complexes to be studied[36]). Moreover, a scanning tunnelling spectroscopy study on TCNE complexation by an adsorbed monolayer of a hexamethylbenzene derivative also found evidence for a tunnelling current increase on TCNE complexation.[40] We therefore investigated the effect of TCNE on the conductance of junctions involving **2**. First, we examined TCNE complexation by 1,4-diethylbenzene (DEB; a surrogate of **2**) in solution. On adding TCNE to DEB in solution in $CH_2Cl_2$, an orange colour was seen due to charge transfer (CT) complex formation. UV-visible spectra were acquired of mixtures of DEB and TCNE as a function of concentrations of donor and acceptor, to enable the equilibrium constant for complexation $K_{DA}$ and the molar extinction coefficient ($\varepsilon$) of the CT peak to be determined using a literature procedure.[36] Two bands, a main peak at 23,800 cm$^{-1}$ and a shoulder at 21,000 cm$^{-1}$ were responsible for the orange colour (Figures S12 and S13) and analysis of the main peak absorbance as a function of donor and acceptor concentrations[36] gave $K_{DA} \approx 1$, $\varepsilon \approx 1000$ dm$^3$ mol$^{-1}$ cm$^{-1}$ for the complex.

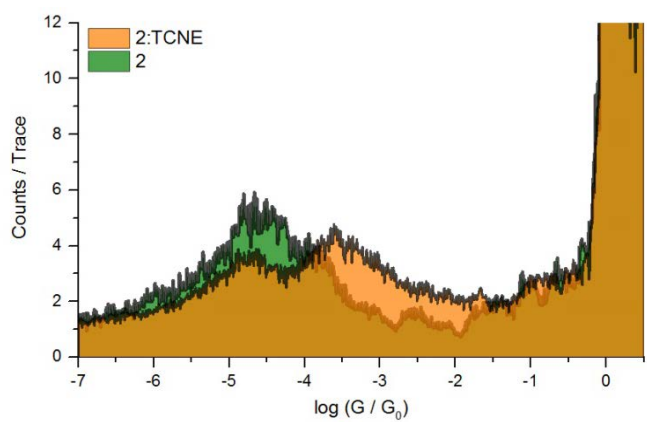

Figure 3 Conductance histogram for STM BJ experiments on self-assembled monolayers of 2 alone (green) and of 2 after treatment with a 2 mM TCNE solution in $CH_2Cl_2$ (orange). Bias voltage 0.3 V.

Once again, we focus here on experimental results obtained using the STM-BJ method for **2** (Figure 3). We were unable to obtain a meaningful conductance histogram for the **2**:TCNE complex in TCB solution in the presence of excess TCNE, probably owing to the low $K_{DA}$ for the complex. We therefore used a modified STM-BJ technique, in which **2** was first adsorbed at high coverage on the Au substrate and the STM-BJ experiment was then conducted in the absence of a liquid medium, using a log-scale pre-amplifier.[37, 41] To enable the formation of the **2**:TCNE complex, the pre-adsorbed **2** was first treated with a 2 mM TCNE solution in $CH_2Cl_2$. It can be seen that in the absence of TCNE, there is a single but very broad conductance histogram peak at $2 \times 10^{-5} G_0$, or 1.5 nS; this is in reasonable agreement with the value previously determined for this molecule using the related *I(s)* technique (0.74±0.24 nS).[38] After treatment of the pre-adsorbed **2** with excess TCNE in $CH_2Cl_2$, two broad peaks are evident in the histogram, at $3 \times 10^{-4} G_0$ and $2 \times 10^{-5} G_0$ (23 nS and 1.5 nS). The latter is identical to the value for free **2**, and we attribute this peak to the formation of some junctions with uncomplexed **2** as a result of the low $K_c$ for the complex; the former is a factor of 15 higher and we attribute this to the formation of junctions with the **2**:TCNE complex.

**Theoretical studies**

To probe the reason for the substantial enhancement of molecular junction conductance upon TCNE charge transfer complexation by **1c** and **2**, we have carried out DFT and NEGF transport calculations. The theory used to describe the conductance enhancement is based on phase-coherent tunnelling, because experimental conductance measurements on single-molecules indicate that this is the dominant transport mechanism up to lengths of order 4 nm. [42, 43]

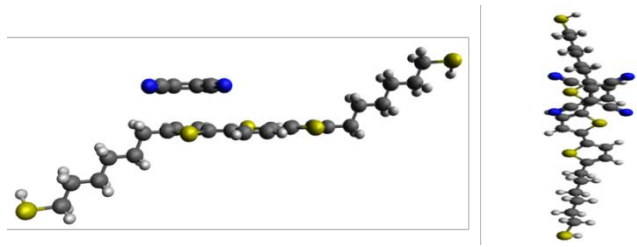

Figure 4 Two views of TCNE bound to the oligothiophene backbone

As a first step we used the DFT code SIESTA[44] (see ESI for technical details) to examine the energy landscape of TCNE bound to **1c**, which possesses several local energy minima. Figure 4 shows the most stable position of TCNE complexed with **1c**. We first used the SMEAGOL code[45] to compute the transmission coefficient $T(E)$ for electrons of energy $E$ passing through the molecular junction from one electrode to the other, for **1c** alone and for the optimally-bound **1c**:TCNE complex. Figure 5 (left panel) shows the resulting plots of $T(E)$ versus $E-E_F^0$, where $E_F^0$ is the 'bare' Fermi energy predicted by DFT. It can be seen that in a new resonance is present in the plot for the **1c**:TCNE complex, and that this is a Fano resonance. A Fano resonance arises when extended states of the molecular junction (including the gold contacts) interact with a bound state.[46, 47] In the examples of Figure 5, the bound state is essentially the LUMO of the TCNE, part-filled by the charge transfer interaction.

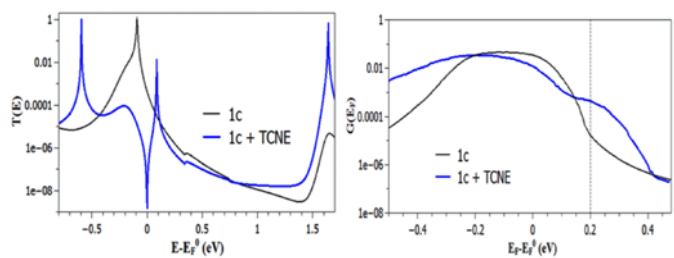

Figure 5 The left panel shows the transmission coefficient $T(E)$ versus $E$ for **1c** in vacuum (black line) and for the optimally-bound **1c**:TCNE complex shown in Figure 4 (blue line). The right panel shows the calculated finite-bias conductance at room temperature (298 K) for **1c** (black line) and the configurationally-averaged, finite-bias conductance for the **1c**:TCNE complex (blue line) as a function of Fermi energy position. The dotted line shows the Fermi energy corresponding to a theoretical conductance enhancement of 27.

In a room temperature experiment, the TCNE is likely to sample a range of positions and orientations relative to the oligothiophene backbone. From an experimental viewpoint, we are interested in the junction conductance and the $I$-$V$ behaviour. We therefore extended the calculations to many different **1c**:TCNE geometries in order to probe the effect of geometry on the position and shape of the Fano resonance (see ESI Figure S16). It is clear from Figure S16 that the position of the Fano resonance around $E-E_F^0$ = 0.1 eV is significantly less dependent on the position and orientation of the TCNE than are the other resonances. A Fano resonance consists of both an anti-resonance (associated with destructive interference) and a nearby resonance (associated with constructive interference).[46] The former is often the focus of attention, because it is a special feature of Fano resonances and does not occur for the more common Breit-Wigner resonances typically observed in plots of $T(E)$, which just show a positive peak. However, the relative importance of destructive or constructive interference depends on the system of interest and in our case, it is the resonance associated with constructive interference that dominates.

Since TCNE is a strong electron acceptor, the position of the orbital responsible for the Fano resonance automatically adjusts to achieve the required filling, and this rationalises why it is relatively independent of the location and orientation of the TCNE relative to the backbone. Furthermore, since a part-filled orbital must be necessarily located near to the Fermi energy $E_F$, it is natural to expect the new resonance to be located near $E_F$ and therefore the transmission coefficient $T(E_F)$, which largely determines the electrical conductance, should be strongly affected. It is well known that DFT cannot accurately predict the true position of the Fermi energy $E_F$ relative to molecular orbitals within a junction and therefore Figure 5 only shows the location of the Fano resonance relative to the 'bare' $E_F^0$. To locate $E_F$, we computed the electrical conductance $G = I/V$ at finite voltage $V$, where

$$I = \left(\frac{2e}{h}\right) \int \int_{-\infty}^{\infty} dE <T(E)> \left(f_+(E) - f_-(E)\right)$$

for a range of values of $E_F$, where $<T(E)>$ is the average of all curves shown in Figure S14 and $f_{+/-}(E)=1/(exp(E-E_F +/- eV/2)/k_BT +1)$ is the Fermi function of the left (+) or right (-) electrode. The right-hand panel of Fig. 5 shows that for a Fermi energy lying in the HOMO-LUMO gap, a conductance enhancement factor of 27 (similar to the measured enhancement), is found at $E_F-E_F^0 = +0.20$ eV (dashed line). This suggests that the experimental Fermi energy $E_F$ is close, but not equal, to the bare DFT value $E_F^0$.

We also carried out similar calculations for **2**. The results are shown in Figure 6, which reveals that complexation with TCNE (optimised geometry shown in Figure S14) again induces a Fano resonance near the Fermi energy leading to an enhancement of the conductance. At $E_F-E_F^0 = +0.20$ eV, the increase is by a factor of 16, close to the experimentally observed value. This shows that the conductance enhancement is a robust signature of the TCNE complexation, rather than a property of a particular donor molecule.

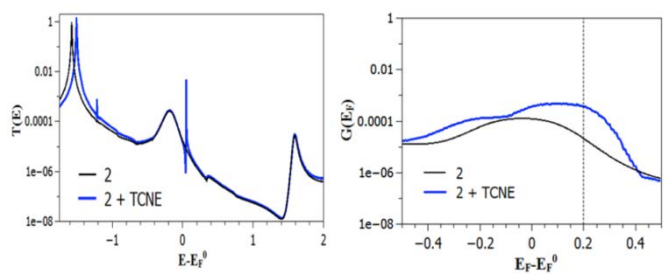

Figure 6 The left panel shows the transmission coefficient *T(E)* versus *E* for **2** in vacuum (black line) and for the optimally-bound **2**:TCNE complex (see Figure S14). The right panel shows the calculated finite-bias conductance at room temperature (298 K) for **2** (black line) and the configurationally-averaged, finite-bias conductance for the **2**:TCNE complex (blue line) as a function of Fermi energy position. The dotted line shows the Fermi energy corresponding to a theoretical conductance enhancement of 16.

## Conclusions

We have shown that charge transfer complex formation boosts the conductance of junctions involving single molecules of **1c** and **2** by up to ca. 20-fold, and that in this case, the explanation is the generation of a Fano resonance in the transmission function near the contact Fermi energy as a consequence of charge transfer complex formation. We are currently examining the relation between the strength of the charge transfer interaction and the conductance enhancement, using other electron acceptors ($I_2$, chloranil, TCNQ). The possible impact of these observations for the future development of single molecule sensing devices is very clear, since they demonstrate that room-temperature quantum interference can be controlled by complexation.

## Experimental

### Syntheses

Molecule 1c was synthesised from 5,5''–dibromo–2,2':5',2''–terthiophene[48] and commercially-available 6–chloro–hex–1–yne by Sonogashira cross-coupling, followed by Pd/C–catalysed hydrogenation, and KI-catalysed nucleophilic displacement using KSAc in acetone; full details and characterisation data are provided in the ESI. Molecule 2 was synthesised as previously described.[38] Conductance measurements

The conductances of molecular junctions were determined using the STM break junction (STM-BJ) method as described in detail in the ESI. Briefly, large Au(111) terraces were first formed on gold on glass or mica substrates by flame annealing.[49] Typically, conductance data were collected by driving a freshly cut Au tip into the substrate at a constant *xy* position (vertical height *z* set to –2 nm) and then withdrawing it at constant speed to *z* +4 nm at 5 nm s$^{-1}$, in the presence of a solution or a monolayer of the target molecular wire. As the tip was pushed into the surface and then retracted, a fresh Au-Au junction was formed, thinned down to a single atom (point contact), and finally broken upon further withdrawal. After the rupture of the junction, a molecule can bridge the tip-substrate gap. The current (*I*) was recorded at a fixed tip-substrate bias (*V*) and conductance *G* is determined by Ohm's law (*G* = *I*/*V*). The feedback loop was subsequently re-engaged and the tip brought to a constant height once more. All the collected conductance-distance traces show plateaus at integer multiples of the quantum of conductance $G_0$, and additional step-like features are observed below 1 $G_0$ in about 30 % of the collected traces. These step-like features are attributed to the conductance of the Au | molecule | Au junction. Thousands of consecutive conductance-distance traces (with no data selection) are compiled in logarithmically binned conductance histograms. These histograms gave a distribution of measured conductance values, with peaks corresponding to the most frequently measured conductance. Measurements were performed in ambient conditions, using 1,2,4-trichlorobenzene solutions (**1c**) or pre-formed self-assembled monolayers (**2**); full details are in the ESI.

**Theory**

Geometry optimizations of molecules **1c** and **2** were performed at the DFT level by the SIESTA code, using LDA[50] to describe the exchange correlation functional, with a double-ζ basis set and norm-conserving pseudopotentials. To find the optimum complex geometry the energies of 500 different TCNE locations with respect to the central units were calculated to find a local minimum. Transport calculations for each of these geometries were then carried out by extending the molecules to include six layers of Au(111) and using an equilibrium Green's function technique (as implemented in the SMEAGOL code) to find the zero-bias transmission coefficient.

## Acknowledgements

We thank EPSRC for funding (grant EP/H035184/1).

## Notes and references